\documentclass[preprints,article,accept,moreauthors,pdftex,10pt,a4paper]{Definitions/mdpi}

\firstpage{1} 
\pubvolume{xx}
\issuenum{1}

\articlenumber{5}
\pubyear{2018}
\copyrightyear{2018}

\newcommand{\Dzero}{{\rm D^0}}
\newcommand{\Dstar}{{\rm D^{*+}}}
\newcommand{\Dplus}{{\rm D^+}}
\newcommand{\pt}{p_{\rm T}}

\newcommand{\sqrtsNN}{\sqrt{s_{\rm \scriptscriptstyle NN}}}
\history{Version January 19, 2019 submitted to Proceedings}

\Title{Multiplicity dependence of heavy-flavour correlations with charged particle and collective effects in p--Pb collisions at $\sqrt{s_\mathrm{NN}}$= 5.02 TeV with ALICE at LHC$^{\dagger}$}

\Author{Marianna Mazzilli for the ALICE Collaboration $^{1,}$$^{2}$\orcidA{}*}

\AuthorNames{Marianna Mazzilli}

\address{%
$^{1}$ \quad Dipartimento Interateneo di Fisica "M. Merlin"\\
$^{2}$ \quad INFN of Bari}

\corres{Correspondence: marianna.mazzilli@ba.infn.it}

\firstnote{Presented at Hot Quarks 2018 - Workshop for young scientists on the physics of ultrarelativistic nucleus-nucleus collisions, Texel, The Netherlands, September 7-14 2018}

\abstract{Azimuthal correlation studies of heavy-flavour particles with charged particles in p--Pb collisions can give an insight into the cold nuclear matter effects \cite{Andronic:2015wma} on heavy-quark production and hadronization into heavy-flavour jets. Multiplicity-dependent measurements of angular correlations of heavy-flavour particles with charged hadrons allow us to investigate the collective behavior of the system and the initial state effects on heavy flavour hadron production. In addition, they can reveal possible modifications of the heavy-quark fragmentation and hadronization at different multiplicities. We present ALICE measurements of azimuthal correlations of prompt D-mesons with charged hadrons as a function of the multiplicity in p--Pb collisions at $\sqrtsNN$ = 5.02TeV. Moreover, the elliptic flow ($v_2$) of heavy-flavour hadron decay electrons in high-multiplicity p--Pb collisions at $\sqrtsNN$= 5.02 TeV, obtained using correlations with charged particles, is reported.}


\begin{document}


\section{Introduction}

Measurements of heavy-flavour production in p--Pb collisions allow to characterize the effects of the presence of a nucleus in the colliding system (cold nuclear matter effects). The analysis of angular correlations between heavy-flavour particles and charged particles is a tool to characterize the heavy-quark fragmentation process, and it is sensitive to their production mechanism in pp collisions. Therefore, differences between the measurements in pp and p--Pb collisions can give an insight into the cold nuclear matter effects on heavy-quark production and hadronization in p--Pb collisions and help to disentangle the effects related to the presence of the Quark-Gluon Plasma in Pb--Pb collisions. 

In addition, multiplicity-dependent studies in p--Pb collisions allow us to investigate the presence of possible collective effects. In recent years, several experiments have observed long-range $v_2$-like angular correlations in high-multiplicity p--A collisions. Their origin is still under debate. Currently, the presence of such structures for heavy-flavour particles is under study. Moreover, it is crucial to study p--A collisions to estabilish possible connections with the behavior of nucleus-nucleus systems (A--A), where collective effects are attributed to the hydrodynamic expansion of the Quark-Gluon Plasma.
To this purpose, azimuthal correlations between D-mesons and charged particles have been studied as a function of centrality in p--Pb collisions at $\sqrtsNN$ = 5.02~~TeV. Analogous studies have been performed for heavy-flavour decay electrons and charged hadrons in p--Pb collisions at $\sqrtsNN$ = 5.02~TeV, as well.

 \section{Azimuthal correlations of D mesons with charged particles}
 Azimuthal correlation distributions of $\Dzero$, $\Dstar$ and $\Dplus$ mesons (trigger particles reconstructed from their hadronic decay channels: $\rm \Dzero\to K^-$$\pi^{+}$ (BR of 3.87$\pm$ 0.05\% \cite{Patrignani:2016xqp}), $\rm D^+\to K^-$$\pi^+$$\pi^+$ (BR of 9.46 $\pm$ 0.24\% \cite{Patrignani:2016xqp}) and $\rm D^{*+}\to D^0$$\pi^+$ (BR of 67.7 $\pm$ 0.5\% \cite{Patrignani:2016xqp}) with charged particles (associated particles) in $|\eta| < $ 1 are evaluated in p--Pb collisions at $\sqrtsNN$ = 5.02 TeV. The analysis has been performed in three different centrality classes defined by the energy deposited in the ZNA detector (Zero-Degree Neutron Calorimeter) in the Pb-going direction: 0-20\%, 20-60\% and 60-100\% \cite{Adam:2014qja}. Several kinematic ranges of the D-meson $\pt$ (3 $< \pt(\mathrm{D}) <$ 24 GeV/$c$) and associated particle $\pt$ (starting from $\pt\mathrm{(assoc)} > $0.3 GeV/$c$) were analysed. 
 
The D-meson candidates were reconstructed by identifying secondary vertices with topologies typical of decays displaced few hundred $\mu$m from the interaction vertex. Geometrical selections on the D-meson decay topology were applied to reduce the combinatorial background. Particle identification was also used to further reduce the background \cite{ALICE-PUBLIC-2017-008}.
The contribution of D-meson combinatorial background is removed by subtracting the correlation distribution evaluated from the sidebands of the D-meson invariant mass distribution.
An event-mixing correction is applied to account for detector inhomogeneities and limited acceptance. Additionally, the distributions are corrected for inefficiencies in the reconstruction and selection of trigger and associated particles. The contribution of D mesons from beauty-hadron decays (ranging from 5 to 10\% depending on the $\pt(\mathrm{D})$) is subtracted, using templates of the angular correlations of feed-down D mesons and charged particles obtained from PYTHIA simulations \cite{Sjostrand:2006za, Sjostrand:2014zea}. 
Since the azimuthal correlation distributions of $\Dzero$, $\Dstar$ and $\Dplus$ mesons are compatible within uncertainties, a weighted average of the three D-meson measurements is performed to reduce the statistical uncertainty. The fully corrected per-trigger azimuthal correlation distributions are fitted with two Gaussian functions, to account for the correlation peaks in the near-side ($\Delta\phi$=0) and away-side ($\Delta\phi$=$\pi$), and a constant (baseline), allowing us to extract quantitative observables such as the near-side associated yield and near-side peak width ($\sigma_{fit, NS}$). 

 \section{Azimuthal correlations of heavy-flavour hadron decay electrons with charged particles}
 
The ALICE Collaboration has measured two-particle correlations between heavy-flavour hadron decay electrons ($e^{HF}$) and charged particles in p--Pb collisions at $\sqrtsNN$= 5.02 TeV \cite{Acharya:2018dxy}. Heavy-flavour hadron decay electrons with transverse momentum in the interval 1.5$< \pt <$ 6 GeV/$c$ and $|\eta| <$ 0.8 were selected by subtracting the contribution of background electrons from the inclusive electron sample. The main background sources arise from photon conversions ($\gamma \to e^{+}e^{-}$) in the beam pipe and in the material of the innermost layers of the Inner Tracking System (ITS) and from Dalitz decays of neutral mesons ($\pi^0 \to \gamma e^{+}e^{-}$ and $\eta \to \gamma e^{+}e^{-}$) \cite{Adam:2015qda}. The analysis was performed in two different multiplicity classes: 0-20\% (high) and 60-100\% (low), obtained using the signal in the scintillator arrays of the V0 detector in the pseudo-rapidity range 2.8$ < \eta < $5.1 (V0A). The correlation distribution for heavy-flavour decay electrons was corrected for the electron and charged particle efficiencies and for the secondary particle contamination. It was also corrected for the limited two-particle acceptance and detector inhomogeneities using the event mixing technique. The two-dimensional correlation distribution was projected onto $\Delta\varphi$ for $|\Delta\eta| <$ 1.2 and divided by the width of the selected $\Delta\eta$ interval. In order to compare the jet-induced peaks from different multiplicity ranges, a "baseline" term, constant in $\Delta\varphi$, was calculated from the weighted average of the three lowest points of the correlation distribution and was subtracted from it. An enhancement of the peaks at $\Delta\varphi$$\approx$0 and $\Delta\varphi$$\approx$$\pi$ can be observed for high-multiplicity collisions. The jet contribution is assumed to be comparable for the two multiplicity classes and removed by subtracting the low-multiplicity from the high-multiplicity distribution. The left panel of Fig.~\ref{fig:v2_henrique} shows the high-multiplicity azimuthal correlation distribution subtracted by the low-multiplicity contribution between heavy-flavour decay electrons with 2 $< \pt <$ 4 GeV/$c$ and charged particles with 0.3$ < \pt < 2$ GeV/$c$.

\section{Results and Conclusions}

\subsection{Azimuthal correlations of D mesons with charged particles}

\begin{figure}
\centering
{\includegraphics[width=0.78\linewidth]{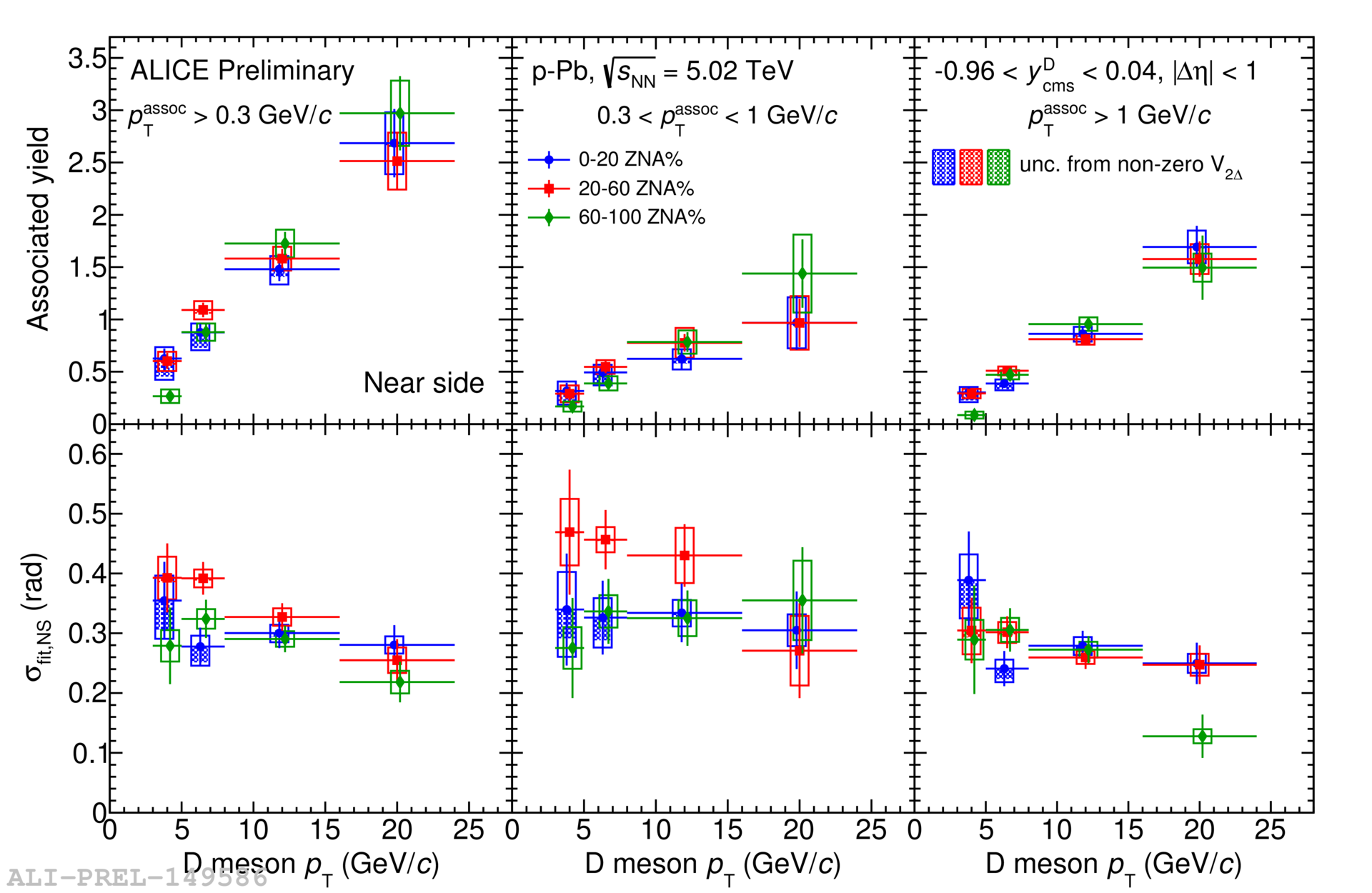}}
 \caption{Comparison of near-side yield and width ($\sigma_{\rm fit,NS}$) in the different centrality classes. Each color is a different centrality class, as described in the legend. The error bars represent the statistical uncertainty, the empty boxes represent the overall systematic uncertainties, the shaded boxes represent the $v_2$ related uncertainties.}
\label{fig:NSresults}
\end{figure}

The comparison of the baseline-subtracted D meson-charged particle azimuthal correlation distributions in p--Pb collisions at $\sqrtsNN$ = 5.02 TeV do not show any significant differences, within uncertainties, among the three centrality classes. The trends of the near-side associated yields and the near-side peak widths extracted in the three centrality classes for p--Pb collisions versus the D-meson $\pt$ are compared in Fig.~\ref{fig:NSresults} for several associated particle $\pt$ ranges. Compatible values of the near-side observables are obtained among the p--Pb collision centralities. No modifications of the near-side peaks due to different heavy-quark fragmentation and hadronization, or the presence of collective effects, are observed. Within the current uncertainties, there is not enough sensitivity to extract a $v_2$ modulation of D-mesons from the azimuthal correlations of D mesons with charged particles.

\subsection{Azimuthal correlations of heavy-flavour hadron decay electrons with charged particles}

\begin{figure}[!htbp]
\centering
{\includegraphics[width=0.49\linewidth]{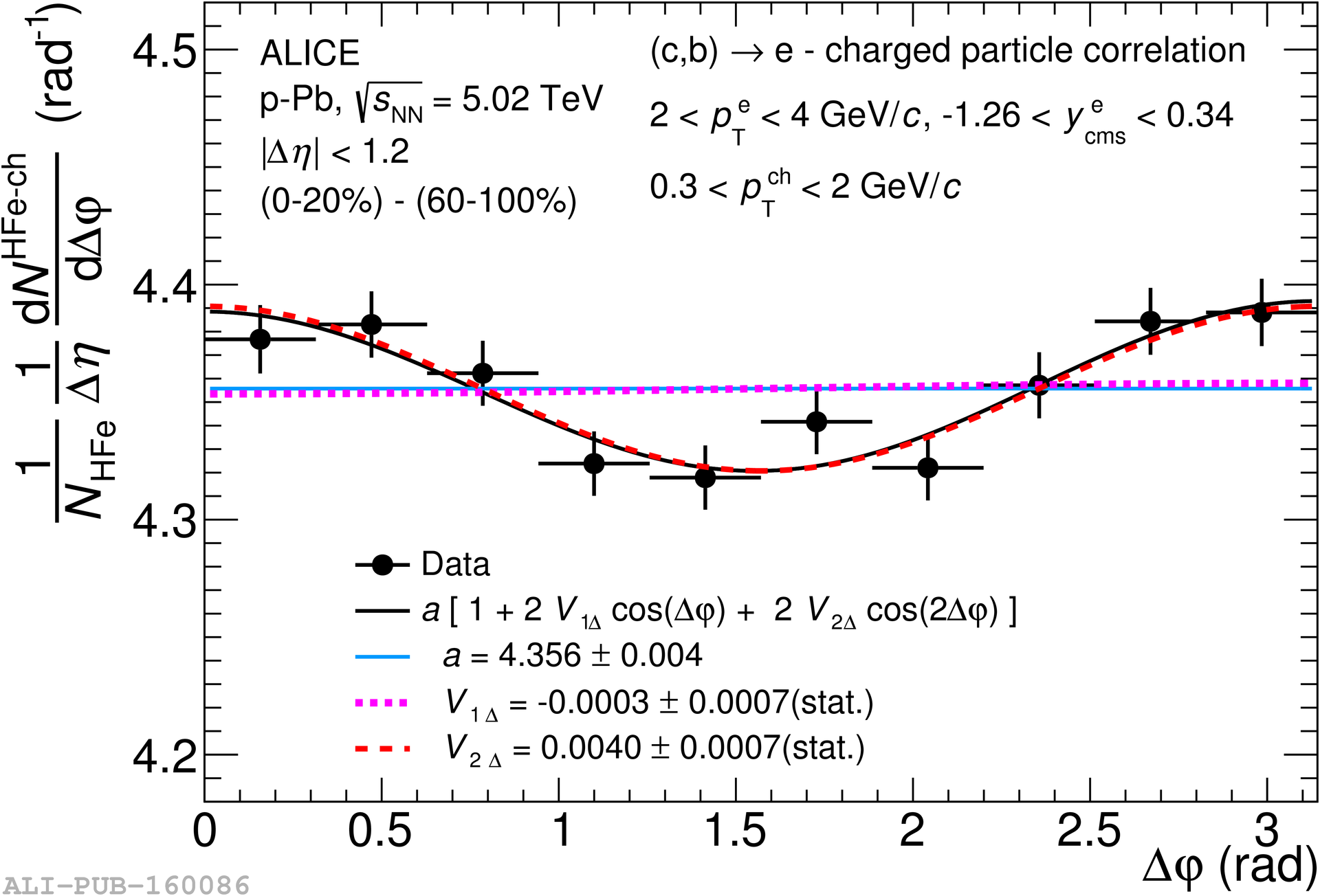}}
{\includegraphics[width=0.49\linewidth]{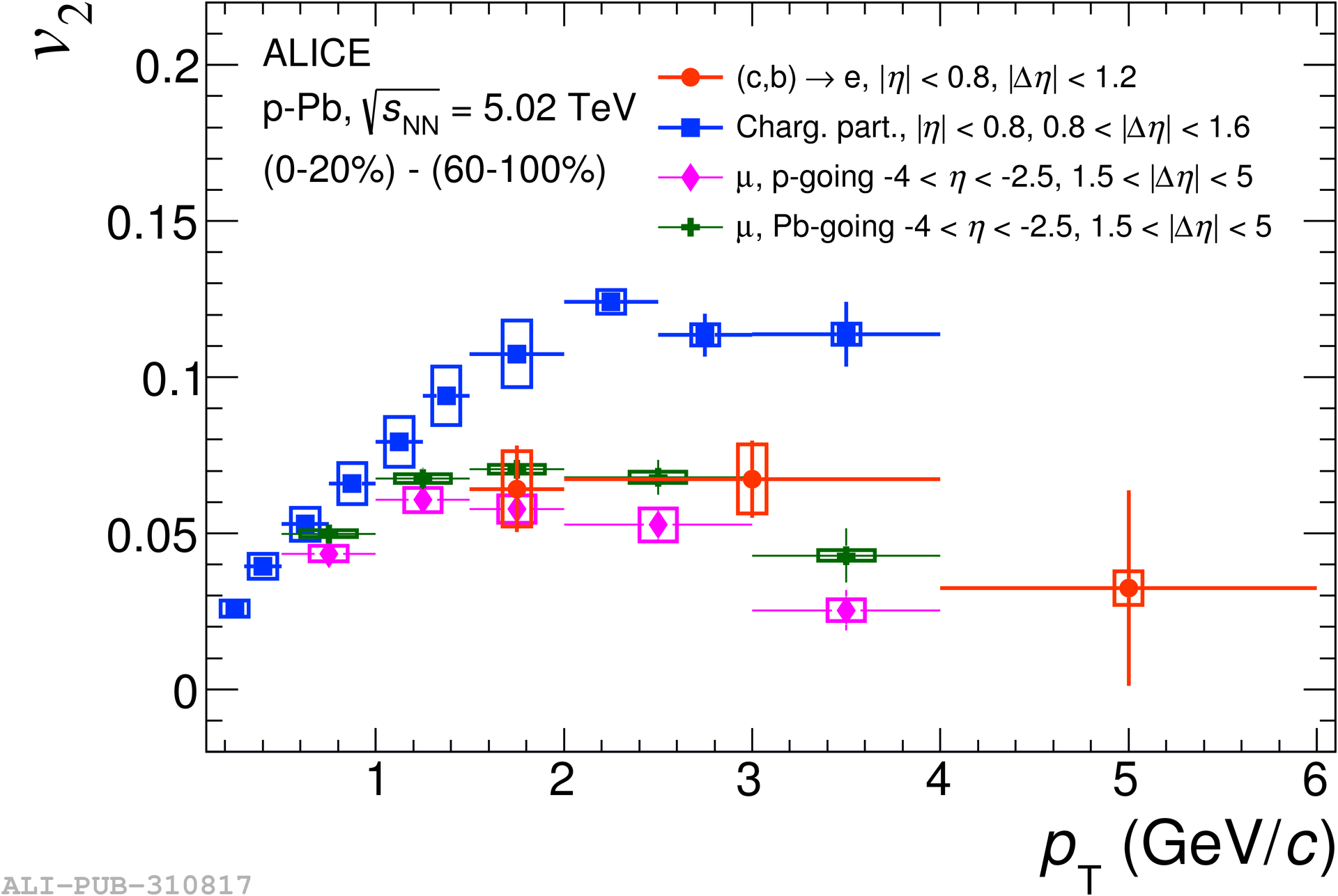}}
\caption{ 
(\textbf{Left}): Azimuthal correlation distribution between heavy-flavour decay electrons and charged particles, for high-multiplicity p--Pb collisions after subtracting the jet contribution based on low-multiplicity collisions. The distribution is shown for 2 $< \pt^e <$ 4 GeV/c and 0.3 $< \pt^{ch} < $2 GeV/c. The fit to the data points and its Fourier decomposition are also shown.
(\textbf{Right}): heavy-flavour decay electron $v_2$ as a function of transverse momentum compared to the $v_2$ of unidentified charged particles \cite{ABELEV:2013wsa} and inclusive muons \cite{Adam:2015bka}. 
 }
\label{fig:v2_henrique}
\end{figure}

After the removal of the jet contribution to the correlation function (left side of Fig.~\ref{fig:v2_henrique}), using the correlation distribution obtained in the low multiplicity class, a $v_2$-like modulation was observed in the high-multiplicity correlation distributions, similarly to what was previously observed for light-flavour di-hadron distributions \cite{Abelev:2012ola}. A Fourier fit to the correlation distributions was used to characterize the modulation and extract the heavy-flavour electron $v_2$. The values of the elliptic-flow coefficient are shown in the right panel of Fig.~\ref{fig:v2_henrique} as a function of the electron transverse momentum. The heavy-flavour decay electron $v_2$ was found to be positive in the transverse momentum interval 1.5 $< \pt^e <$ 4 GeV/$c$ in high-multiplicity events with a significance of 5.1 $\sigma$. The values of heavy-flavour decay electron $v_2$ are generally lower, though still comparable, than those measured for charged particles dominated by light-flavour hadrons. An interpretation of this comparison is anyway not immediate since the transverse-momentum distribution of heavy-flavour hadrons from which the electrons originate is considerably broader than that of light-flavour hadrons \cite{ABELEV:2013wsa} while also having a larger average value. The $v_2$ of heavy-flavour decay electrons has a similar trend of that of inclusive muons, measured by ALICE at forward and backward rapidity and dominated by heavy-flavour decay muons for $\pt >$ 2 GeV/$c$ \cite{Adam:2015bka}. Drawing conclusions from this comparison is not straightforward because of the different cold-nuclear-matter effects affecting heavy flavours in different rapidity ranges. This measurement complements previous measurements for light-flavour hadrons and provides new information on the behaviour of heavy-flavour hadrons to understand the azimuthal anisotropies observed in small collision systems.

\reftitle{References}


\end{document}